\documentclass[twocolumn,aps,prl]{revtex4}
\usepackage{chapterbib}
\usepackage{graphicx}
\usepackage{dcolumn}
\usepackage{bm}
\begin{document}
\title{\textit{Ab initio} study of ladder-type polymers polythiophene and polypyrrole}

\author{Simon $\text{Pesant}^1$, Paul $\text{Boulanger}^1$, Michel $\text{C\^ot\'e}^1$ and Matthias $\text{Ernzerhof}^2$}

\affiliation{$~^1$D\'epartement de physique et Groupe de recherche en physique et technologie des couches minces
(GCM),
Universit\'e de Montr\'eal, C. P. 6128 Succursale Centre-ville,
Montr\'eal (Qu\'ebec) H3C 3J7, Canada}

\affiliation{$~^2$D\'epartement de chimie
Universit\'e de Montr\'eal, C. P. 6128 Succursale Centre-ville,
Montr\'eal (Qu\'ebec) H3C 3J7, Canada}
\email{Michel.Cote@umontreal.ca}

\begin{abstract}
This article presents an \textit{ab initio} study of four polymers, polythiophene, polypyrrole, ladder-type polythiophene, and ladder-type polypyrrole. Upon an analysis of the variation of the band gap when comparing the unconstrained and the ladder-type polymers, a discrepancy was found  between the thiophene and the pyrrole polymer families. For polythiophene, the ladder-type polymer has a larger gap than the unconstrained polymer whereas the opposite is found for the pyrrole polymers. The structural properties  and the charge densities using the Bader charge analysis of these four compounds are investigated. The different band gap behaviors in thiophene and pyrrole polymers can be explained in terms of the competition between the bond length alternation and the effect of the charge density in the carbon backbone.    
\end{abstract}
\maketitle
 \section{Introduction}
 
Conjugated polymers raise much interest because of their intriguing electronic properties and their foreseen technological applications. They have many advantages when compared to inorganic semiconductors such as their easy processing, their tunable optical gap and their  favorable structural properties. Their electronic properties are mainly due to their delocalized $\pi$-electrons along their carbon backbone.
Polyanniline\cite{naturepani}\cite{panidoped}, polypyrrole\cite{ppybattery} and polythiophene are examples of polymers with promising future developments. In particular, polythiophene and its derivatives were used in several applications such as numeric display device\cite{applithioscience}, surface light emitting diode(SLED) \cite{applithiosynthmet} and light emitting diode(LED)\cite{applithioadvmat}. 

One particular class of polymer that is of present interest is the ladder-type polymers. These have additional bonds as compared to other polymers which link the neighboring monomers rigidly together eliminating the possible dihedral degree of freedom. These polymers are known generally to exhibit small band gaps, due partly to their planar configurations which maximize the alignment of the $\pi$ orbitals. Moreover, these ladder-type polymers have the potential to exhibit very high intrachain mobility\cite{mobility_ladder}.

Recently, Oyaizu and al.~\cite{ladderpolythiosynth} reported the first synthesis of a ladder-type polythiophene(\textbf{LPT}) and published details about its electronic structure that they have characterized both theoretically and experimentally. Using a parameterized Hartree-Fock model (PM5), they found a band gap reduction in \textbf{LPT} as compared to polythiophene(\textbf{PT}); they also obtained indirect experimental evidences of this gap reduction. They relate this behavior to a reduction of the bond length alternation observed in their calculations when the backbone of the LPT is compared to PT. This argument is similar to the one used in polyacetylene which explains the band gap and the dimerization of the atomic structure.

In this article, a pseudopotential density-functional theory study of the electronic properties of four polymers, the polythiophene(\textbf{PT}), ladder polythiophene(\textbf{LPT}), polypyrrole(\textbf{PPy}),  ladder polypyrrole (\textbf{LPPy}) is presented. The results indicate a decrease of the band gap for the ladder-type version of polypyrrole, however, on the contrary, the ladder-type polythiophene  present a larger band gap as compared to \textbf{PT}. This difference between these two systems is examined in detail in the remaining of this article.

 \section{Computational Methods}

The results reported in this article were computed within the framework of the density-functional theory(DFT) as implemented in two codes that use different basis set to represent the electronic degree of freedom. The first is the Abinit package\cite{abinitcode} which represents the electronic states with a plane-wave basis set, useful in the pseudopotential formulation of periodic structures. The exchange-correlation energy was calculated both in the local density approximation(LDA) using the Teter-Pade parametrization which reproduces the Ceperley-Alder data\cite{LDA} and in the Perdew-Burke-Ernzerhof(PBE) generalized-gradient approximation\cite{PBE}. The pseudopotentials were generated with the Trouiller-Martins scheme\cite{trouillermartins} and their portability was fully tested. Numerical convergence of the total energy within 1 mHa/atom was reached for a number of plane-waves in the basis set corresponding to a kinetic energy cutoff of 35 Ha and a sampling of the Brillouin zone of 8 k-points on a shifted grid.

The Abinit package is a solid-state oriented code, which implicitly generates Born-von Karman periodic boundary conditions in all directions. This implies that within Abinit, the electronic structure of a single polymer is not directly computed, but rather the electronic structure of a uniform array of polymers is simulated. The unwanted interaction between neighbor polymers can easily be minimized by increasing the space between polymers which however, also augments the computational efforts. This approach has the advantage that the electronic degrees of freedom are represented by an orthogonal basis set which covers all space equivalently and that can be check for completeness simply by increasing the number of plane-waves, e.i. the kinetic energy cutoff mentioned above.  

The charge densities presented in Fig.~\ref{atomstructure} was obtained using Bader charge density analysis~\cite{Bader}. To get converged results, a basis that includes plane-waves up to a kinetic energy of 100 Ha was required to get the charge density on a very fine grid in real space.

Furthermore, to compare the validity of the functionals used and of the results for isolated polymers, the Gaussian 03 code\cite{g03} was employed. This code is oriented towards molecules and represents the wave functions with a gaussian basis set. The basis used was 6-31G\cite{basis1, basis2, basis3, basis4, basis5, basis6, basis7, basis8, basis9}
and this choice will be explained later. Within Gaussian 03, periodicity can be imposed only in the directions of interest. In the present case, only one translation vector is specified to obtain truly one isolated polymer. The exchange and correlation functionals used were again the LDA and the PBE to compare the results for a single polymer. As can be seen in Table~\ref{tab:gap}, the band gap values obtained with Abinit and Gaussian for the \textbf{PPy} and \textbf{LPPy} are similar, but for the \textbf{PT} and \textbf{LPT} a small difference is noted. This variation can be explained, in part, by the different pseudopotentials used in the two simulations. Nevertheless, the band gap variations reported in Table~\ref{tab:var_gap} for these polymers are comparable and differ only by \mbox{0.1 eV} between the Abinit and the Gaussian results. To investigate the validity of the bond length alternation obtained with LDA and PBE, the calculations with the B3LYP functional\cite{b3lyp} was also employed which includes exact exchange. This functional is implemented in the Gaussian 03 code but not in the Abinit package. Finally, the Gaussian 03 code used an equivalent of 96 k-points to integrate over the Brillouin zone.

The polymers have been initialized in a coplanar conformation and have been fully optimized. All the polymers are considered uncharged. 
Although the relaxation of the atomic structure was not constrained the coplanar symmetry was never broken during the optimization process, both with the Abinit and the Gaussian codes. This is consistent with the expectation of an uncharged $\pi$-conjugated system and with the results obtained for the oligomer versions of these polymers\cite{hutchison_ratner}. 
The treatment of periodic systems within the Gaussian code is a fairly new functionality. Unfortunately, the B3LYP functional is not well suited for periodic systems. 
The long-range part of the exact exchange in this functional causes convergence problems which are correlated to the basis set. It was found that the 6-31G  basis set was the largest that will converge for all the polymers. This correlation is simple to understand, reducing the basis set effectively reduces the long-range overlaps between wave functions making the exact exchange part of the B3LYP functional more manageable. 
Fortunately, these long-range effects have only little influences on the properties studied and the 6-31G basis set is sufficiant the treat the present electronic systems.


\section{Results and Discussion}
\subsection{Thiophene based polymers}

\begin{figure}
\includegraphics[height=0.8 \linewidth]{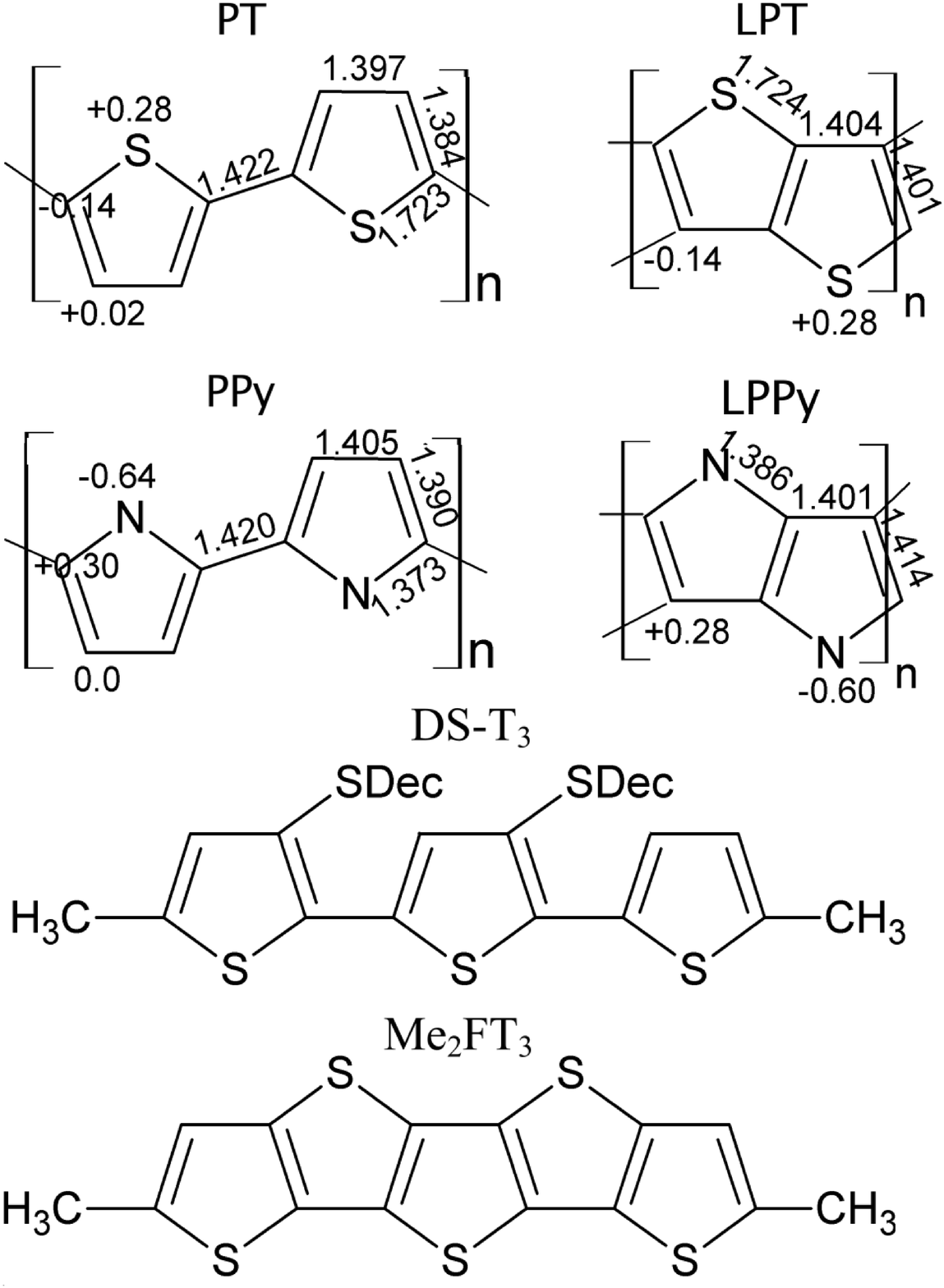}
\caption{\label{atomstructure}Atomic structure of the polymers and oligomers. The charges and bond length obtained within LDA are indicated on the atomic structures of the polymers.}
\end{figure}

The first synthesis of \textbf{LPT} was reported by Oyaizu and al.\cite{ladderpolythiosynth}. They explored the electronic structure of this polymer with indirect experiments and by performing calculations using the PM5 level of theory, which uses empirical data to parametrize the two center integrals within the Hartree-Fock theory. This method explicitly includes exchange contribution, but partially includes correlation effects through the empirical fitting of the electronic properties. Within this method, the atomic structure of the \textbf{PT} is characterized by a clear bond length alternation in the carbon backbone. The bond opposite to the sulfur atom within the thiophene monomer is practically the same length as the intermonomer bond, being only 0.011 {\AA} shorter, whereas the other C-C bonds within the thiophene is 0.073 {\AA} shorter. The bond alternation obtained is therefore about $\delta r \sim 0.06$ {\AA} for \textbf{PT}.  However, with the \textbf{LPT} atomic structure, there are only two inequivalent C-C bonds and they are found to be roughly of same length resulting in a bond alternation of just $\delta r = 0.005$ {\AA} for this polymer. All carbon atoms are equivalent in the \textbf{LPT}, whereas there are two inequivalent carbon atoms in the PT.

The authors then discuss the electronic structure obtained for the optimized structure. The band gap for both polymers corresponds to a direct transition at k=0. The calculated band gap for \textbf{PT} is found to be 6.45 eV while for \textbf{LPT} it is found to be 5.92 eV. They argue that the missing correlation effects should only result in a constant shift of the unoccupied electronic levels, which should reduce the band gap of \textbf{PT} to the experimental value of 2.1 eV. Applying the same shift leads to a band gap of 1.5 - 1.6 eV for \textbf{LPT}. The reduction of the band gap between the two polymers is then explained using the reduced bond alternation by a tight-binding-like argument as it is done in the polyacetylene case. This result is compared with experiment via UV absorbtion measurements, photoluminescence measurements and electrochemical properties (redox reactions) of thin films. The main difficulty of these procedures are that \textbf{LPT} is insoluble, forming a opaque film, and cannot be easily functionalized. This means that direct measurement of the UV absorption and the photoluminescence of \textbf{LPT} in solution is not possible. Results for the oligomers and the closest soluble precursor of both polymers are presented and suggest a band gap diminution. They also report results on both non-ladder oligomers and ladder oligomer. Fig.~\ref{atomstructure} shows the non-ladder trimer oligothiophene(\textbf{DS-}$\mathbf{T_3}$) and its ladder version($\mathbf{Me_2FT_3}$). The UV absorption spectra and the photoluminescence intensity maximums are both found to be red-shifted for the ladder version of the oligomer. 

\begin{figure}
\includegraphics[height=0.8 \linewidth]{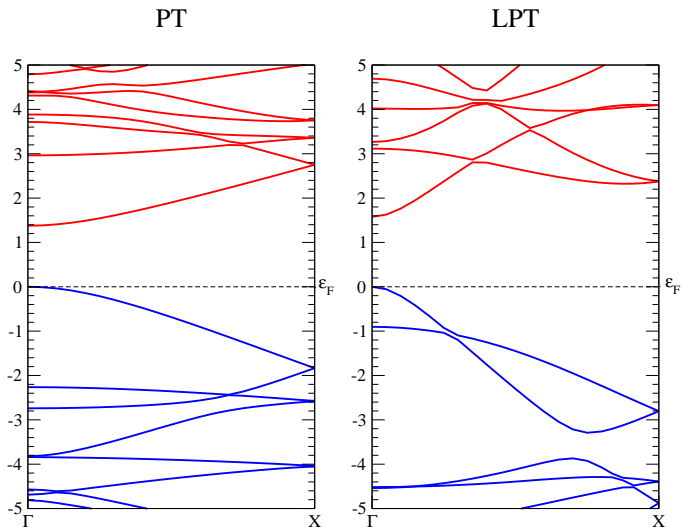}
\caption{\label{ptgroup} LDA band structure of \textbf{PT} and \textbf{LPT}}
\end{figure}

The results obtained in the present study are in contrast with the results presented above, the band gap of \textbf{LPT} is found to be 0.18 to 0.33 eV larger than \textbf{PT} depending on the functional used, as shown in Tables~\ref{tab:gap} and \ref{tab:var_gap}. The LDA band structure of these two polymers are depicted in Fig.~\ref{ptgroup}.  The band gap obtained with B3LYP is much more consistent with the experimental value of 2.1 eV for \textbf{PT}\cite{gap_pt}.
After optimization of the structures with the B3LYP (LDA) functional, the bond length alternation is found to be $\delta r \sim 0.04 (0.03)$ {\AA} for \textbf{PT} and $\delta r = 0.016 (0.003)$ {\AA} for \textbf{LPT}. The B3LYP results show that the bond alternation is increased compared to LDA and PBE functionals, which as already been seen in the case of trans-polyacetylene, thiophene based oligomers\cite{hutchison_ratner} and polyyne oligomers\cite{imamura_aoki}. 

\begin{table}[!h]
\begin{tabular}{|c|c|c|c|c|c|}
\cline{2-6}
\multicolumn{1}{c}{} & \multicolumn{2}{|c|}{Abinit}  &  \multicolumn{3}{|c|}{Gaussian}   \\ \hline
polymer & LDA & PBE & LDA & PBE & B3LYP  \\ \hline
\textbf{PT} & 1.38 & 1.05 & 0.892 & 0.865 & 1.852  \\ \hline
\textbf{LPT} & 1.58 & 1.51 & 1.148 & 1.091 & 2.031  \\ \hline
\textbf{PPy} & 1.76 & 1.80 & 1.760 & 1.722 & 2.826   \\ \hline
\textbf{LPPy} & 1.61 & 1.55 & 1.680 & 1.644 & 2.595  \\ \hline
\end{tabular}
\caption{\label{tab:gap}Calculated band gaps for different functionals of the selected polymers given in eV. }
\end{table}

\begin{table}[!ht]
\begin{tabular}{|c|c|c|c|c|c|}
\cline{2-6}
\multicolumn{1}{c}{} & \multicolumn{2}{|c|}{Abinit}  &  \multicolumn{3}{|c|}{Gaussian}  \\ \hline
polymer & LDA & PBE & LDA  & PBE & B3LYP \\ \hline
\textbf{LPT} - \textbf{PT} & 0.335 & 0.309 & 0.256 & 0.226 & 0.179 \\ \hline
\textbf{LPPy} - \textbf{PPy} & -0.182 & -0.24 & -0.080 & -0.078 & -0.231  \\ \hline
\end{tabular}
\caption{\label{tab:var_gap}Variation of the band gaps of \mbox{Table~\ref{tab:gap}} given in eV.}
 \end{table}
 
The primary concern will be the variation of the band gap between a polymer and its ladder-type equivalent.The net result of the increased bond alternation in B3LYP is an almost uniform increase in the band gaps of all the polymers. Therefore, the difference between B3LYP and LDA functionals will not be an issue since both give the same variational behaviour, as illustrated in Table~\ref{tab:var_gap} for clarity. 
 

A survey of the literature reveals that the standard behavior as been a decrease of the band gap in the ladder-type form of a polymer. It is the case for benchmark polymers like polyfluorene and polycarbazole\cite{briere_cote}. The band gap reduction in these polymers can be readily understood by the structural changes incurred during the ladder transformation. The unconstrained polymers are not coplanar: they have a non-zero value for the dihedral angles between monomers, varying between 26 and 27 degrees which is a result of the electrostatic repulsion between the hydrogen atoms or other side groups. This causes a misalignment of the $\pi$ orbitals that decreases the dispersion of their bands resulting in an increase of the band gap. On the other hand, the planarity is enforced by the additional bonds in the ladder-type polymers. The $\pi$ orbitals interaction between monomers is maximized which favors a delocalisation of the wave functions and cause an increase in the dispersion of these levels, hence, a reduction of the band gap.

Generally, the ladder-type polymers exhibit smaller band gaps then their unconstrained counterpart because they are planar. However, this cannot be the case for polythiophene, since the original polymer is already coplanar. The variation of the band gap must therefore be linked to more subtle changes of the geometry. 

As noted by Chung and al.\cite{chung}, the absorption properties of \textbf{PT} closely resemble that of \emph{trans}-polyacetylene. The backbone consists of \emph{trans} segments linked through a \emph{cis}-like unit which implies that its electronic properties will lie between the two structures. A simple nearest-neighbour tight-binding model of the $\pi$-electrons of \emph{trans}-polyacetylene shows that the bond alternation will be critical in the determination of the band gap. All the carbons atoms (with their corresponding hydrogen) in this structure are equivalent. Hence, if the bond lengths are equal, the matrix elements corresponding to tight-binding hopping parameters are also all equal. This leads to a closure of the band gap at the Brillouin zone edge. As mentioned before, in the case of \textbf{PT} the carbon atoms in the backbone are not all equivalent, which suggest that even if the bond alternation is suppressed, the band gap should decrease but might not completely vanish. 

The key assumption in the preceding argument was the decrease in the bond alternation while preserving the \emph{trans}-polyacetylene like conformation. However, in the ladder-type polymers, the backbone has changed as compared to the non-ladder version and they now resemble more closely to that of  \emph{cis}-polyacetylene. In the  \emph{cis} case, the carbons atoms are again all equivalent, but the hopping matrix elements are not. This is a subtle consequence of the geometry. A close consideration of the position of the hydrogen atoms reveals the change in their orientation between two consecutive pairs of carbon atoms. Thus, even with equal bond lengths there is no symmetry enforcing the values of the hopping terms to be equals and the band gap at the zone edge does not vanish. 

The present results can now be understood with these arguments. The ladder-type polymer does decrease the bond alternation in the backbone, but in the new conformation, this does not reduce the gap. A perfect example of this is the 1.148 eV band gap with no bond alternation found in \textbf{LPT} with the LDA formalism obtained with the Gaussian code. The main reason for this band gap variation is due to the changes in the atomic structure between the non-ladder and the ladder types polymers. In the non-ladder type, the polymer is made of \emph{trans}- and \emph{cis}-segments whereas the ladder-type is entirely made of \emph{cis}-segments. Care must be taken when comparisons are made with experiments, because the results stated for the polymers will not necessarily be correspond to the oligomers. The tight-binding model and the band representation, breaks down without the Born-von Karman periodic boundary condition which means that a direct generalization to the oligomer cases cannot be made. Even the electronic structure of the precursors may be dominated by others effects related to the side-chains, as explained later in the case of \textbf{PPy}. 

 To address the experimental results directly, the HOMO-LUMO separation of both oligomers showed in Fig.~\ref{atomstructure} using the B3LYP functional were calculated.  A value of 3.25 eV  is obtained for  \textbf{DS-}$\mathbf{T_3}$ and 3.53 eV for $\mathbf{Me_2FT_3}$. For simplicity, the side-chains in  \textbf{DS-}$\mathbf{T_3}$ were not explicitly included since they are not expected to influence the HOMO or LUMO energies.  Even in the case of the oligomers the band gap is found to increase in the ladder-type conformation which is contrary to the reported behavior observed in the optical spectra. The differences may be the consequence of a solvent interactions or of exciton effects that are not fully taken in consideration in the present calculations. 

\subsection{Pyrrole based polymers} 

\begin{figure}
\includegraphics[height=0.8 \linewidth]{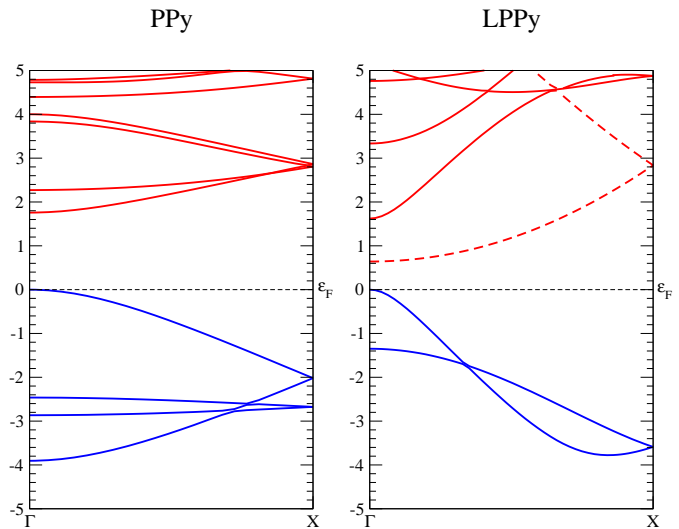}
\caption{\label{ppygroup} LDA band structure of \textbf{PPy} and \textbf{LPPy}. The dashed line represents the nearly free electron states(NFE).}
\end{figure}

Interestingly, the simple tight-binding model seems to fail for the polypyrrole family. All the DFT calculations shows that the band gap of \textbf{LPPy} is lower than the one of \textbf{PPy}, as seen in Table ~\ref{tab:gap} and Fig.~\ref{ppygroup}. For the \textbf{LPPy}, the dashed line is related to a nearly free electron states (NFE). This state is characterized by orbitals located outside the polymer, such states are also seen in graphite\cite{NFE}. The gap for \textbf{LPPy} reported in Table~\ref{tab:gap} neglects the NFE states.   Nevertheless, the  change in the backbone symmetry should generate an increase in the band gap even for the \textbf{LPPy}. This suggests that another competing phenomenon must be taken into account to explain the band gap variation of all ladder-type polymers. 

In order to find  the difference between the pyrrole and thiophene polymer families, the band structures for their fully relaxed backbones only was computed. These correspond to the cis-polyacetylene for \textbf{LPT} and \textbf{LPPy} while for \textbf{PT} and \textbf{PPy} it is a mix between cis and trans-polyacetylene. The impact on the band gap of adding a sulfur or nitrogen atoms to these backbones can then be studied. Within LDA, the gap for cis-polyacetylene is 0.5 eV, compared to the 1.58 eV and 1.61 eV for \textbf{LPT} and \textbf{LPPy} respectively. This band gap increase is similar for the two polymers, which demonstrates that the effect of the added atoms is important but also that it doesn't depend strongly on its nature. 
The bond alternations in these ladder-type polymers are similar and thus their band gaps have roughly the same value.

For the non-ladder polymers, the backbone presents a LDA gap of 0.8 eV, in contrast with 1.38 eV for \textbf{PT} and 1.76 eV for \textbf{PPy}. A difference between adding an nitrogen or sulfur atom is now observed. The $\sim$0.4 eV increase of the band gap between \textbf{PPy} and \textbf{PT} can't be accounted for by the variation of the bond length alternation as seen in Fig.~\ref{atomstructure}. Thus, the intrinsic properties of the added atoms must be considered to explain the difference in band gaps. The larger electronegativity of nitrogen as compared to sulfur causes a greater electron attraction in its local environment. This implies that its nearest neighbor atoms should be positively charged. At a smaller scale, the same phenomenon is found in \textbf{PT}. As mentionned before, these polymers are composed of two types of carbon atoms. In the case of \textbf{PPy}, the charge disparity between the two types of carbon is 0.3e, in contrast with \textbf{PT} which is 0.17e. This results in different hopping matrix element parameters in a tight-binding model. Since the backbone behave more like a trans-polyacetylene, there will be a larger opening of the gap for the \textbf{PPy} than for the \textbf{PT}. 
Hence, to fully explain the electronic properties of these polymers it is necessary to examine the charges distribution in the polymer's backbone.

In the case of the ladder polymers, the sulfur or nitrogen atoms still attract electrons in their surroundings, as seen in the Bader charge values reported on Fig.~\ref{atomstructure}. But, in the \textbf{LPT} and \textbf{LPPy}, all carbon atoms are identical by symmetry. Therefore, all the charges are equal. This implies that the charge alternation is not a variable in our tight-binding model and would not contribute in the band gap value. This explain why it is not necessary to take into consideration the charge variation in those ladder-type polymers to determine their band gaps. 


\section{Conclusion}

To conclude, the present study fully explains the behavior of band gaps for two groups of polymers. It was shown that the band gap of the \textbf{LPT} was larger compared to \textbf{PT}. For the pyrrole family, the results found were opposite, the band gap of \textbf{LPPy} being lower than for \textbf{PPy}. To account for these results, a competition between two phenomenons is presented, the bond alternation of the atomic structure and the charge of the carbon atoms. These two, depending which one dominates, can give rise to an increase or decrease of the band gap of the polymer in consideration. 
 
\section{Acknowledgments}
This work was supported by grants from the FQRNT and NSERC. The computational ressources were provided by the R\'eseau qu\'eb\'ecois de calculs de haute performance (RQCHP). 

\bibliography{polymereponte.bib}
\end{document}